\documentclass[3p,times]{elsarticle}

\usepackage{ecrc}

\volume{00}

\firstpage{1}

\journalname{Nuclear Physics A}

\runauth{}

\jid{npa}

\jnltitlelogo{Nuclear Physics A}

\usepackage{amssymb}

\biboptions{square,comma,numbers,sort&compress}

\usepackage[figuresright]{rotating}

\begin{document}

\begin{frontmatter}



\dochead{Proceeding contribution for Hard Probes 2013}

\title{The Search for Gluon Saturation in pA Collisions}


\author{Bo-Wen Xiao}

\address{Key Laboratory of Quark and Lepton Physics (MOE) and Institute
of Particle Physics, Central China Normal University, Wuhan 430079, China}

\begin{abstract}
In this talk, I reviewed the recent theoretical progress towards the exploration of the gluon saturation phenomenon in $pA$ collisions at both RHIC and the LHC.
Two important pillars of this exploration are the single inclusive forward hadron productions and forward dihadron correlations, which have both been computed up to one-loop order within the small-$x$ factorization formalism. Furthermore, detailed numerical study for the single inclusive hadron production at next-to-leading order provides us good description of the low transverse momentum data measured in $dAu$ collisions at RHIC with significantly reduced uncertainty. In addition, the one-loop calculation for dihadron productions in $pA$ collisions shows that there is a new type of large logarithm, which is known as the Sudakov factor, arising in the back-to-back correlation limit. 
\end{abstract}


\end{frontmatter}


\section{Introduction}
\label{intro}
Small-x physics provides us the description of dense parton densities at high energy limit when the longitudinal momentum fraction $x$ of partons with respect to parent hadron is small. Furthermore, it predicts the onset of the gluon saturation phenomenon\cite{Gribov:1984tu, Mueller:1985wy, McLerran:1993ni} as a result of nonlinear QCD dynamics when the gluon density is high. Due to Bremsstrahlung gluon radiations, which are enhanced in the small-$x$ regime, gluon density is expected to rise rapidly as $x$ decreases. This effect is known to be governed by the famous Balitskii, Fadin, Kuraev, and Lipatov (BFKL) evolution equation\cite{Balitsky:1978ic}, which results in a resummation of $\alpha_s \ln\frac{1}{x}$. In addition, when too many
gluons are present in a confined hadron with fixed size, they start to overlap and recombine. The balance between the gluon radiation and recombination yields the so-called gluon saturation. To include the recombination effect, a nonlinear term in the evolution equation was introduced\cite{Gribov:1984tu, Mueller:1985wy}. The non-linear extension of BFKL evolution equation was derived by Balitsky and Kovchegov, and therefore normally referred as BK evolution equation\cite{Balitsky:1995ub, Kovchegov:1999yj}. In practice, one can define the so-called saturation scale to separate the nonlinear saturated dense regime from the linear dilute regime. The core ingredients of saturation physics are the multiple interactions and small-$x$ evolutions. It is conceivable that the multiple interactions become important when parton densities, which is driven by the BK equation, are of order $\mathcal{O}(\frac{1}{\alpha_s})$. It seems obvious that the parton saturation phenomenon is an inevitable consequence of QCD dynamics in high energy scatterings. The remaining important tasks are to improve small-$x$ physics calculation with more precise predictions and look for the signatures of gluon saturation in high energy experiments both at RHIC and the LHC.

There have been a lot of progress in terms of the exploration of the saturation physics in $pA$ collisions.
Two of the key efforts focus on the studies of the single inclusive forward hadron production and dihadron productions in $pA$ collisions. The advantage of $pA$ collisions is that the projectile proton can be viewed as a relatively dilute object which probes the dense nuclear target during the scattering without dealing with the complication of the factorization issue\cite{Collins:2007nk}. The complicated multiple interactions between the probe and nucleus target can be shuffled into the modified parton distributions\cite{Bomhof:2006dp,Xiao:2010sp, Xiao:2010sa}. In fact, one can demonstrate and use an effective hybrid factorization for forward hadron productions in $pA$ collisions for both of the single\cite{Chirilli:2011km, Chirilli:2012jd} and dihadron\cite{Dominguez:2010xd, Dominguez:2011wm, Mueller:2012uf, Mueller:2013wwa} forward productions. In addition, to use the single inclusive hadron production as a precise probe of saturation phenomenon, a program called Saturation physics at One Loop Order (SOLO)\cite{Stasto:2013cha}, is developed. SOLO is a numerical program which systematically includes almost all of the next-to-leading (NLO) corrections for single hadron productions in $pA$ collisions, and it is the first important step towards the precision test of saturation beyond the leading logarithmic approximation.  

\section{Inclusive Forward Hadron Productions in $pA$ collisions}
In terms of the effective factorization \cite{Dumitru:2002qt, Dumitru:2005gt}, inclusive forward hadron productions in $pA$ collisions,
\begin{equation}
p+A\to h (y, p_\perp)+X \ ,\label{pA}
\end{equation}
can be viewed as a large $x$ parton fragments into hadron $h$ at rapidity $y$ with transverse momentum $p_\perp$ after going through target nucleus and multiply interacting with the dense gluonic matter inside the target. The cross section associated with this process can be written as follows
\begin{eqnarray}
\frac{d\sigma}{dyd^2p_\perp} =\int xf_a(x)\otimes D_a(z) \otimes \mathcal{F}^{x_g}_a(k_\perp) \otimes \mathcal{H}^{(0)} +\frac{\alpha_s }{2\pi} \int xf_a(x)\otimes D_b(z) \otimes \mathcal{F}_{(N)ab}^{x_g}\otimes \mathcal{H}_{ab}^{(1)},
\end{eqnarray}
where the first term corresponds to the leading order (LO) contribution, while the second term represents the NLO contributions computed from one-loop diagrams with $a,b=q, g$. The full expression for the above cross section is available in Ref.~\cite{Chirilli:2011km, Chirilli:2012jd}. According to kinematics, it is straightforward to find $x=\frac{k_\perp}{\sqrt{s}}e^y$ and $x_g=\frac{k_\perp}{\sqrt{s}}e^{-y}$ with $zk_\perp=p_\perp$ and $z$ being the longitudinal momentum fraction of produced hadron with respect to its original parton. The LO contribution together with some running coupling effects has been studied quite extensively\cite{Albacete:2010bs, Levin:2010dw, Fujii:2011fh, Albacete:2012xq, Lappi:2013zma}.

Since we are focusing on the hadron productions at forward rapidity $y$, $x$ is so large that partons from the proton side can be described by collinear parton distributions. Therefore, the transverse momentum of produced hadron at LO is solely from the transverse momentum of the small-$x$ gluon from the nucleus target. At NLO, hard splittings can also contribute to the transverse momentum of the produced hadron\cite{Altinoluk:2011qy}. This is the main reason that the NLO correction can be important since it opens up new channels and brings additional source to the transverse momentum of the produced hadron. 

NLO corrections can be computed from the one-loop diagrams, which normally contains various types of divergences \cite{Chirilli:2011km, Chirilli:2012jd}. For total cross section, divergences often cancels between real and virtual graphs, the uncanceled divergences are the collinear divergences which are associated with the long distance physics and can be absorbed into the redefinition of collinear parton distributions $xf(x)$ or/and fragmentation functions $D(z)$. This procedure is commonly known as the renormalization of parton distributions and fragmentation functions, which yields the well-known DGLAP evolution equation. In this case, since we also measure the transverse momenta of produced hadrons, the cancellation between real and virtual diagrams are even less complete. This introduces a new type of divergences, namely the so-called rapidity divergence. Through detailed computation, we demonstrate that this rapidity divergence can be removed through the renormalization of the transverse momentum dependent gluon distribution $\mathcal{F}^{x_g}(k_\perp)$ from the target nucleus, and the corresponding evolution equation is identical to the well-known BK equation. 

At the end of the day, we can show that all the divergences at one-loop order can either cancel between real and virtual graphs, or can be renormalized into the DGLAP or BK evolution equations. The rest of the one-loop contribution contains no divergence, therefore can be put into the hard factors $\mathcal{H}$. Furthermore, by solving the evolution equations, we can resum all the corresponding large logarithms $(\alpha_s L)^n$ up to all order in $n$ with $L$ representing large log. For example, by solving the LO DGLAP equation, we can resum all the leading logarithms of the type $\alpha_s \ln\mu^2 $. In addition, by using NLO parton distributions which are solutions of NLO DGLAP equation, we can further include the NLO $\alpha_s$ correction from parton distributions. In principle, in order to achieve the complete NLO $\alpha_s$ accuracy for inclusive hadron productions, we should include not only the NLO hard factor $\frac{\alpha_s }{2\pi}  \mathcal{H}_{ab}^{(1)}$, but also use the NLO solutions to all the relevant evolution equations together with running coupling corrections. 

In the recently developed SOLO program\cite{Stasto:2013cha}, we have been incorporate all the NLO corrections except for the NLO BK evolution due to some technical difficulty. SOLO gives decent agreement with the experimental data\cite{Arsene:2004ux, Adams:2006uz} from RHIC for forward rapidity hadron production when $p_\perp < Q_s$ with significantly reduced renormalization scale dependence. Thus, it is more precise and reliable than the LO results in this region. However, for $p_\perp >Q_s$, one observes that the cross section drastically drops to negative region. It was a surprise at first sight, however, this should be normal since fixed order calculations beyond LO are not guaranteed to be positive. In addition, the saturation formalism is not expected to work at high $p_\perp$ where perturbative QCD calculation dominates. This may indicate that the small-$x$ calculation breaks down in this region where parton densities are dilute therefore multiple interactions are no longer important. One possible solution to this negative cross section problem is to include the exact kinematics presented in this conference\cite{Beuf:2014uia}. One can also simply use perturbative QCD calculation for the large $p_\perp$ region with the exact kinematics or some sort of resummation and match towards the low $p_\perp$ region with the small-$x$ calculation\cite{ta}.

\section{Forward Di-hadron Productions in $pA$ collisions}

\begin{figure}[htb]
\begin{center}
\includegraphics[width=6cm]{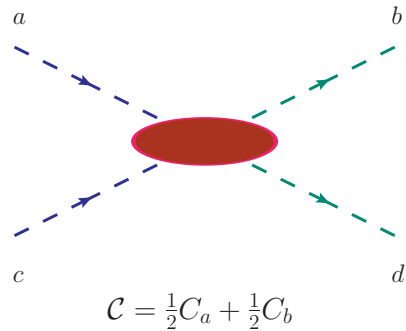}
\end{center}
\vspace*{8pt}
\caption{Illustration of the dijet production processes ($a+b\to c+d$) in general and the corresponding colour factor of the Sudakov factor.\label{f1}}
\end{figure}

Forward dihadron productions in $pA$ collisions have been reckoned as one of most interesting probes which are directly sensitive to the gluon saturation effects in the nucleus target. Assuming that leading $p_\perp$ hadrons can be viewed as surrogates of jets, the dominant contribution of this process comes from dijet productions as schematically shown in Fig.~\ref{f1}. 

The physical picture at partonic level is that a parton coming from the proton projectile splitting into two partons before or after going through the dense nucleus medium and eventually producing two back-to-back jets at approximately the same forward rapidity $y$. Although the transverse momentum of each jet $P_{i\perp}$ can be large, the transverse momentum imbalance $q_\perp=|P_{1\perp}+P_{2\perp}|$ of these two jets remains relatively small, and mainly comes from the small-$x$ gluon with transverse momentum $q_\perp$ originated from the target nucleus, assuming that the intrinsic transverse momentum of partons inside the projectile proton is negligible. By measuring the azimuthal angle $\phi$ correlation of these two jets, especially the away side correlation, where the azimuthal angle difference $\Delta \phi \simeq \pi$, one can probe the typical transverse momentum of small-$x$ gluon, thereby extract information about the saturation effects inside target nucleus. 

The transverse momentum of small-$x$ gluon $q_\perp$ typically is of the same order of the saturation momentum $Q_s$, which is a scale characterising the strength of the saturation effect. If the saturation effect is small, $Q_s$ is small, therefore, we expect to see a strong peak for back-to-back correlations in dijet productions. On the other hand, if the saturation effect becomes strong, we anticipate that the momentum imbalance between these two jets gets large, therefore the away side ($\Delta \phi \simeq \pi$ where $P_\perp\simeq P_{i\perp}\gg q_\perp$) peak of dihadron gets suppressed. Measuring this process at forward rapidity allows us to interact with gluons with smaller $x$, which gives more room for the BK evolution, hence yields larger saturation momentum $Q_s$. 

The quantitative feature of this suppression was first predicted by Ref.~\cite{Marquet:2007vb}. There have been some compelling experimental evidences \cite{Braidot:2010ig, Adare:2011sc, Li:2012bn} on the suppression of the back-to-back dihadron correlations at forward rapidity in $dAu$ collisions at RHIC. Afterwards, based on the saturation physics framework, more elaborated theoretical\cite{Dominguez:2010xd, Dominguez:2011wm} and numerical\cite{Albacete:2010pg, Stasto:2011ru, Lappi:2012nh} studies of this dihadron correlation find very good agreement with the experimental data. So far, all the phenomenological studies of the di-hadron correlations in $pA$ collisions are still based on the LO calculation. To reliably investigate this process and make predictions for $pPb$ collisions at the LHC, one should go beyond LO by including the one-loop contributions. In fact, we have found that there are new types of large logarithms arising from one-loop calculations, e.g., the Sudakov double logarithms $\alpha_s \ln^2 \frac{P_\perp^2}{q_\perp^2}$.

\begin{figure}[htb]
\begin{center}
\includegraphics[width=5cm]{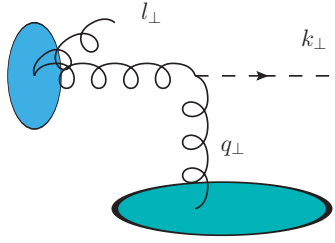}
\end{center}
\vspace*{8pt}
\caption{Illustration of one gluon emission for Higgs production in $pA$ collisions.\label{f2}}
\end{figure}

Through a detailed rigorous one-loop calculation\cite{Mueller:2012uf, Mueller:2013wwa} for Higgs productions in $pA$ collisions as illustrated in Fig.~\ref{f2}, after showing that all the divergences either cancel between real and virtual graphs or can be renormalised into corresponding evolution equations, we manage to demonstrate that small-$x$ resummation can be separated from the resummation of the Sudakov double logarithms (also known as the Collins-Soper-Sterman (CSS) resummation\cite{Collins:1984kg}) at least at the level of the leading logarithm approximation. It is conceivable that there could be a unified description of the small-$x$ evolution and the CSS evolution for the unintegrated gluon distributions involved in a hard scattering such as Higgs productions at high energy. The Sudakov double logarithm is generated because of incomplete real and virtual
graph cancellation when the phase space for additional gluon emission is limited. As a comparison, it is well known that there is normally no large logarithms for inclusive observables due to complete cancellation between real and virtual diagrams. As shown in Fig.~\ref{f2}, because Higgs is heavy and colourless, the gluon cloud (the blue blob), which originally belongs to the wave function of the horizontal incoming gluon, is forced to be produced with transverse momentum $l_\perp$. In principle, one can integrate over the full phase space of the radiated gluon. On the other hand, in order to produce a Higgs particle with fixed transverse momentum $k_\perp$, the phase space of the real emission is highly limited ($l_\perp$ can not be too large than $k_\perp$) while there is no such restriction on the virtual graphs. Therefore, virtual contributions shall dominate over the real contribution and give rise to the so-called Sudakov suppression factor. 

Furthermore, we have studied the Sudakov factor for dijet productions in pA collisions. For this kind of processes, suppose we can write down a generic expression as
\begin{equation}
\frac{d\sigma}{dy_1dy_2 dP_\perp^2  d^2q_\perp}\propto H(P_\perp^2)\int d^2x_\perp d^2y_\perp e^{iq_\perp\cdot (x_\perp-y_\perp)}
{\widetilde{W}}_{x_A}(x_\perp,y_\perp) \ , \label{gen}
\end{equation}
where $H(P_\perp)$ corresponds to the hard factor only depending on the 
hard momentum scale $P_\perp$ and the rapidities of the two jets ($y_1$, $y_2$), $\widetilde{W}_{x_A}$ 
is defined as the Wilson line correlators associated with the un-integrated gluon distributions
of the target nucleus. After the Sudakov resummation, Eq.~(\ref{gen}) can be modified as
\begin{equation}
\frac{d\sigma}{dy_1 dy_2 dP_\perp^2 d^2q_\perp} \propto H(P_\perp^2)\int d^2x_\perp d^2y_\perp e^{iq_\perp\cdot R_\perp}
e^{-{\cal S}_{sud}(P_\perp, R_\perp)}{\widetilde{W}}_{x_A}(x_\perp,y_\perp) \ ,
\end{equation}
where the Sudakov factor can be written as
\begin{equation}
{\cal S}_{sud}=\frac{\alpha_s}{\pi}{\mathcal C}\int_{c_0^2/R_\perp^2}^{P_\perp^2}\frac{d\mu^2}{\mu^2}\ln\frac{P_\perp^2}{\mu^2}  \ .
\end{equation}
where $c_0=2e^{-\gamma_E}$ and $R_\perp=x_\perp-y_\perp$.
The ${\mathcal C}$ coefficient is the colour factor for the double logarithm, which can be obtained through a detailed analysis of gluon radiation
at one-loop order. We found that ${\mathcal C}=\frac{1}{2}C_a+\frac{1}{2}C_b$ which only depends on the colour flows of the incoming partons a and b (blue dash lines) as shown in Fig.~\ref{f1}. For example, we find that $C_a=C_F$ for incoming quarks, $C_a=C_A$ for incoming gluons. This is due to the fact that the outgoing final state dijets (the green dashed lines as shown in Fig.~\ref{f1}) are very close to each others in the coordinate space and their colours are highly correlated. Since the Sudakov double logarithms come from the soft and collinear region of the radiated gluon, only the incoming partons contribute while the soft and collinear gluon contribution from the outgoing partons cancels. 

\section{Conclusion}

In conclusion, for the last few years, a lot of progress in the saturation formalism has been made towards better theoretical understanding and more precise phenomenological description for various observables, especially the single inclusive hadron and dihadron productions in forward $pA$ collisions. The numerical implementation of the NLO corrections for single inclusive hadron productions makes the important precision test of saturation physics possible. As another possible smoking gun measurement, the suppression of the forward back-to-back dihadron (dijet) correlations in $dAu$ collisions at RHIC has provided us the striking evidence for gluon saturation. With the implementation of the Sudakov factor for this process and further development, it is conceivable that more complete and reliable prediction shall be available soon. 





\bibliographystyle{elsarticle-num}



\end{document}